# Quantifying the effects of nickel on Earth's inner-core nucleation

Jiahui Zhai[1#], Liangrui Wei[1#], Chen Gao[1], Yang Sun[1*]
[1]*Department of Physics, Xiamen University, Xiamen 361005, China*
(Dated: August 28, 2026)

**Abstract**
The formation of Earth's solid inner core marks a major transition in the thermal and chemical evolution of the deep Earth, yet its origin remains paradoxical, as initial nucleation appears to require unrealistically large undercooling in the outer core. Here, we use atomistic simulations to quantify how Ni affects this process under inner-core conditions. While Fe-Ni alloys preserve strong thermodynamic competition between the hcp and bcc phases, the bcc phase consistently forms smaller critical nuclei and has lower nucleation barriers than hcp. Increasing Ni content in the melts further lowers the nucleation barrier and shortens the nucleation waiting time. Local chemical fluctuations also strongly affect the macroscopic nucleation rate. Combining these effects, bcc nucleation in $Fe_{80}Ni_{20}$ reaches about 250 K of undercooling, approaching geophysical constraints. We demonstrate that Ni enrichment, bcc nucleation, and chemical heterogeneity substantially narrow the inner-core nucleation paradox.

## 1. Introduction

The solidification of the Earth's inner core (IC) provides a key energy source to sustain the geodynamo through the release of latent heat and the partition of light elements [1,2]. Although seismology and mineral physics constrain many present-day properties of the inner core [3–7], the mechanism that initiated its solidification remains an open question. Thermal evolution models usually assume that nucleation begins once the core temperature falls below. the melting point [8]. However, while the solid becomes more stable than the liquid below the melting temperature, nucleation still requires sufficient undercooling to overcome the barrier associated with forming the solid-liquid interface. Calculations with classical nucleation theory (CNT) suggest that the iron nucleation requires substantial undercooling far larger than the plausible value that core can reach with the secular cooling, leading to the inner core nucleation paradox [9,10].

One way to explore the paradox is to investigate intrinsic nucleation pathways in pure Fe. Molecular dynamics (MD) simulations show that Fe crystallization can proceed through a kinetically favored two-step pathway mediated by a metastable body-centered cubic (bcc) phase, whose nucleation requires less undercooling than that of the stable hexagonal close-packed (hcp) phase [11]. This scenario is consistent with Ostwald's step rule, which predicts that a metastable polymorph can serve as a kinetic gateway, as observed in many materials [12,13]. In metallic systems, bcc often has a lower solid-liquid interfacial energy than close-packed structures [14,15]. Although hcp remains the stable phase of Fe at IC pressures [16–18], *ab initio* and machine-learning simulations indicate that bcc becomes energetically competitive near melting [19–26]. Recent experiments also suggest the transient bcc formation during thermal cycling close to inner-core conditions [27]. These results suggest that bcc-mediated two-step nucleation is likely relevant to the onset of inner-core solidification [28].

In addition to pure Fe, the core is expected to contain Ni and several light elements [29]. Previous studies have considered how Fe alloying with light elements of S, Si, O, and C modifies its nucleation without assuming which phase nucleates [30–33]. These studies showed that while elements such as C and O can lower the nucleation barrier, their addition simultaneously depresses the melting point. In contrast, Si and S severely hinder the nucleation process, significantly increasing the required undercooling. However, the effect of Ni on nucleation rates has not been quantified, despite its abundance in the core. Geochemical models suggest a mean core concentration of ~ 5 wt.% Ni [6,34], while the higher Ni contents of up to 15 wt. % observed in iron meteorites, which represent fragments of differentiated planetesimal cores, motivated examining a wider range of Fe-Ni compositions in core studies [35–38].

Ni has been shown to play a critical role in influencing the thermodynamic properties of Fe under core conditions. Experiments of Fe-Ni alloy suggests possible formation of bcc phase near the core conditions [37,39]. *Ab initio* simulations suggested that Ni can increase the bcc stability in the Fe–Ni alloys [40–42]. It is also shown that Ni has a much higher melting point than Fe under IC conditions [43], unlike the light elements that mostly only reduce the Fe's melting point [31]. Together, these data imply that Ni may influence both phase competition and nucleation kinetics in a way that is distinct from light-element alloying.

In this work, we quantify how Ni affects the nucleation of Fe-Ni alloys under Earth's inner-core conditions. We combine the thermodynamic calculations with nucleation simulations to determine the critical nucleus sizes and the free-energy barriers and for both bcc and hcp nucleation in the Fe-Ni alloy. We further analyze spontaneous compositional fluctuations in the Fe-Ni liquid and evaluate how they modify the effective nucleation rate. We aim to connect phase stability, nucleation pathway selection, and local chemical fluctuations during the initial crystallization of the Earth's inner core.

---

[#]These authors contributed equally.
[*]Email: yangsun@xmu.edu.cn

## 2. Methods

We calculated Gibbs free energies of liquid, hcp, and bcc Fe and Ni using Gibbs-Helmholtz integration from the melting point obtained by solid-liquid coexistence simulations [11,44]. Alloy free energies were computed with a regular solution model [42]. Critical nucleus sizes were determined with the persistent-embryo method (PEM), which facilitates nucleation simulations with an adaptive potential without biasing the critical region [45]. Nucleation rates were obtained from steady-state nucleation theory using atomic attachment rates computed by the iso-configurational ensembles [46]. Additional simulation details and equations are provided in the Supplementary Information.

All molecular dynamics simulations were performed in LAMMPS [47] with an embedded-atom Fe-Ni potential refined from our previous work [42] to better reproduce Fe's *ab initio* free energy and Fe-Ni mixing enthalpy (see Supplementary Information). Simulations used the NPT ensemble with the Nosé-Hoover thermostat and barostat [48]. The MD timestep was 1 *fs*. For the PEM simulations, the system size was set to ~ 53,000 atoms. Local structures were identified with cluster alignment [49] and polyhedral template matching [50] and thresholds from the equal-mislabeling criterion [51]. During nucleation simulations, a small fraction of interfacial atoms occasionally exhibited fcc-like order within the hcp phase, which are classified as stacking-fault structures of the hcp phase.

## 3. Results

### 3.1 Fe-Ni Phase Diagram and Free Energy

We first compare the free energy and phase diagram of the Fe-Ni system calculated using the present EAM potential with previous DFT results [42]. Figure 1(a) shows the comparison of relative Gibbs free energy potential referenced to the liquid phase between the current EAM potential and DFT calculations. Near the melting points, the deviations in $\Delta G^{S-L}$ are only a few meV/atom, and the corresponding shifts in the endmember melting temperatures are less than 70 K. Thus, the present potential accurately describes the thermodynamics of both liquid and solid Fe and Ni endmembers. In Fig. 1(b), the EAM potential predicts a liquid mixing enthalpy that agrees well with the DFT results. The mixing enthalpy of the hcp phase is also well reproduced, with differences of less than 5 meV/atom compared to the DFT data. It also correctly captures the relative ordering of the mixing enthalpies of the bcc, hcp, and liquid phases. However, the potential does not reproduce the negative mixing enthalpy of the bcc phase at ~5 at.% Ni and its rapid increase at 10–15 at.% Ni. This trend of mixing enthalpy is a deviation from regular-solution behavior and indicates anomalous ordering in the Fe–Ni alloy, analogous to the inversion of short-range order reported in Fe–Cr alloys [52,53]. Reproducing such a low-concentration anomaly in Fe–Cr required an explicitly concentration-dependent EAM potential [52], which is substantially more complicated than the present form. These differences in free energy cause the Fe-Ni phase diagram in Fig. 1(c) to exhibit a larger bcc stability region and higher liquidus temperature by ~100 K in the 5-8 at.% Ni range compared to the *ab initio* phase diagram in [42]. Nevertheless, the key features of the phase diagram are preserved. In particular, the liquidus temperature increases with increasing Ni composition, indicating that Ni stabilizes the solid phases relative to the liquid under IC conditions. The phase diagram also retains the coexistence of the bcc and hcp phases near the liquidus, capturing the competition between these phases with increasing Ni content. The potential further shows good agreement with DFT results for the liquid structures of Fe, Ni, and Fe–Ni alloys in Supplementary Material Fig. S1. We therefore employ the present potential to investigate the nucleation behavior of Fe–Ni alloys.

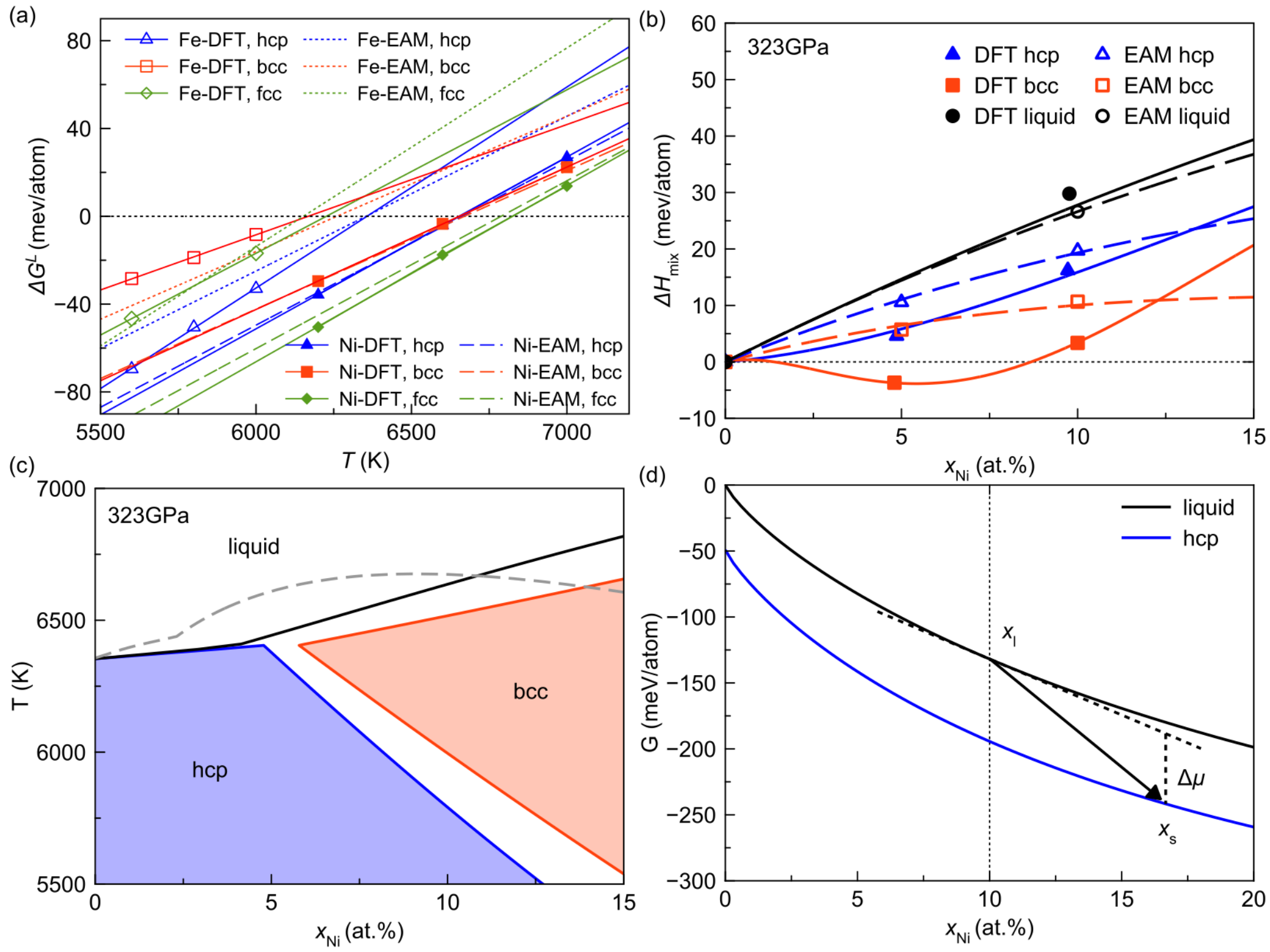

**FIG. 1.** Gibbs free energy and phase diagram of the Fe-Ni alloy at 323 GPa. (a) Relative Gibbs free energy referenced to the liquid phase. The DFT data are taken from [42]. (b) Mixing enthalpies, $\Delta H_{\text{mix}}$, of hcp, bcc, and liquid Fe-Ni solutions at 6000 K. The lines denote fits to the regular solution model. (c) Fe-Ni phase diagram. The dashed line shows the liquidus curve from DFT calculation [42]. (d) Calculation of the nucleation driving force ($\Delta\mu$) for a nucleus with composition $x_S$ emerging from a liquid with composition $x_l$. The dashed line is tangent to the liquid free-energy curve.

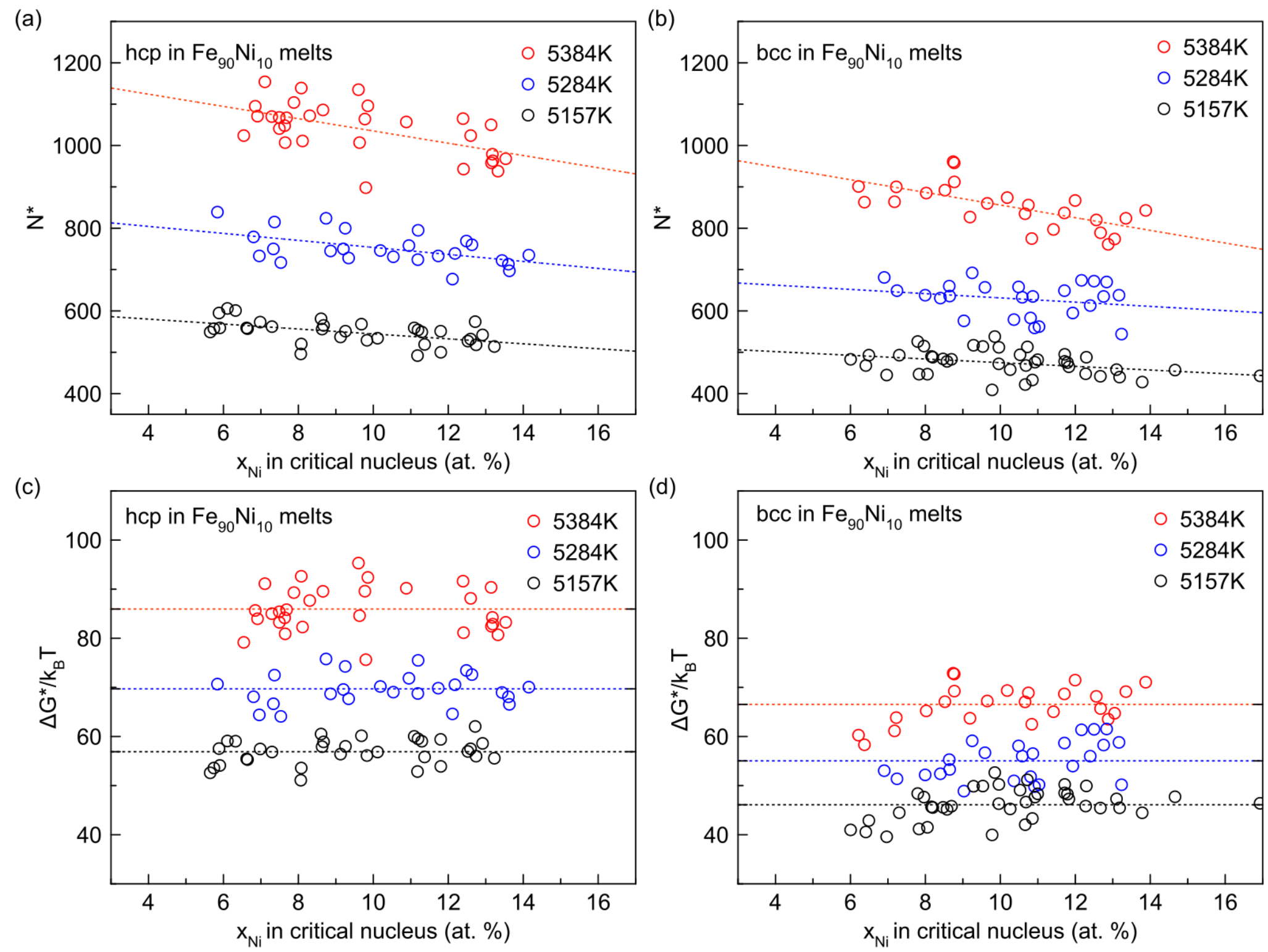

**FIG. 2.** Critical nucleus size and nucleation barrier of $Fe_{90}Ni_{10}$ liquid at 323 GPa. Size and composition of the critical nuclei for the (a) hcp and (b) bcc phases. The dotted lines indicate linear fits. Nucleation free-energy barriers for the (c) hcp and (d) bcc phases.

The free energy data computed via the EAM potential allows us to compute the nucleation driving force, $\Delta\mu$. For pure Fe, the nucleation driving force is the same to the Gibbs free energy between solid and liquid, i.e., $\Delta\mu_{Fe} = \Delta G_{Fe}^{s-l}$ [11,30]. In a liquid solution, the nucleus can have a different composition with the liquid. Thus, the nucleation driving force should be defined as the chemical potential change associated with the phase transition as illustrated in Fig. 1(d), expressed as

$$\Delta\mu(x_s, x_l) = G^s(x_s) - \left[ G^l(x_l) + (x_s - x_l)\frac{\partial G^l(x)}{\partial x}\bigg|_{x_l} \right] \quad (1)$$

where $G^l(x_l)$ and $G^s(x_s)$ are the Gibbs free energies of the liquid and solid at compositions of $x_l$ and $x_s$, respectively [54].

### 3.2 Nucleation in Fe-Ni Solutions

To probe the nucleation kinetics, we perform PEM simulations for $Fe_{90}Ni_{10}$ liquid at 5157, 5284, and 5384 K. At this composition, the melting temperatures of the hcp and bcc phases are 6484 and 6531 K, respectively, based on the phase diagram in Fig. 1(c). The initial Ni content of the embryo is set between 5 and 15 at.%. Representative PEM trajectories are provided in Supplementary Material Figs. S2 and S3 for the bcc and hcp phases, respectively. Unlike the single critical nucleus observed in pure Fe and Ni systems [11,45], the critical nuclei formed in the binary liquid solution can exhibit significantly different sizes and compositions while still reaching the critical region. Complete statistics on the size and composition of the critical nuclei for the hcp and bcc phases obtained from the PEM simulations are shown in Fig. 2(a) and (b). The critical nuclei exhibit a wide range of Ni compositions and sizes, with compositions centered around that of the liquid. At a fixed liquid composition, nuclei with higher Ni content tend to have smaller critical sizes. The nucleus size also decreases systematically with decreasing temperature. Comparison between the bcc and hcp results further shows that the bcc phase consistently has a smaller critical nucleus size than the hcp phase at the same temperature.

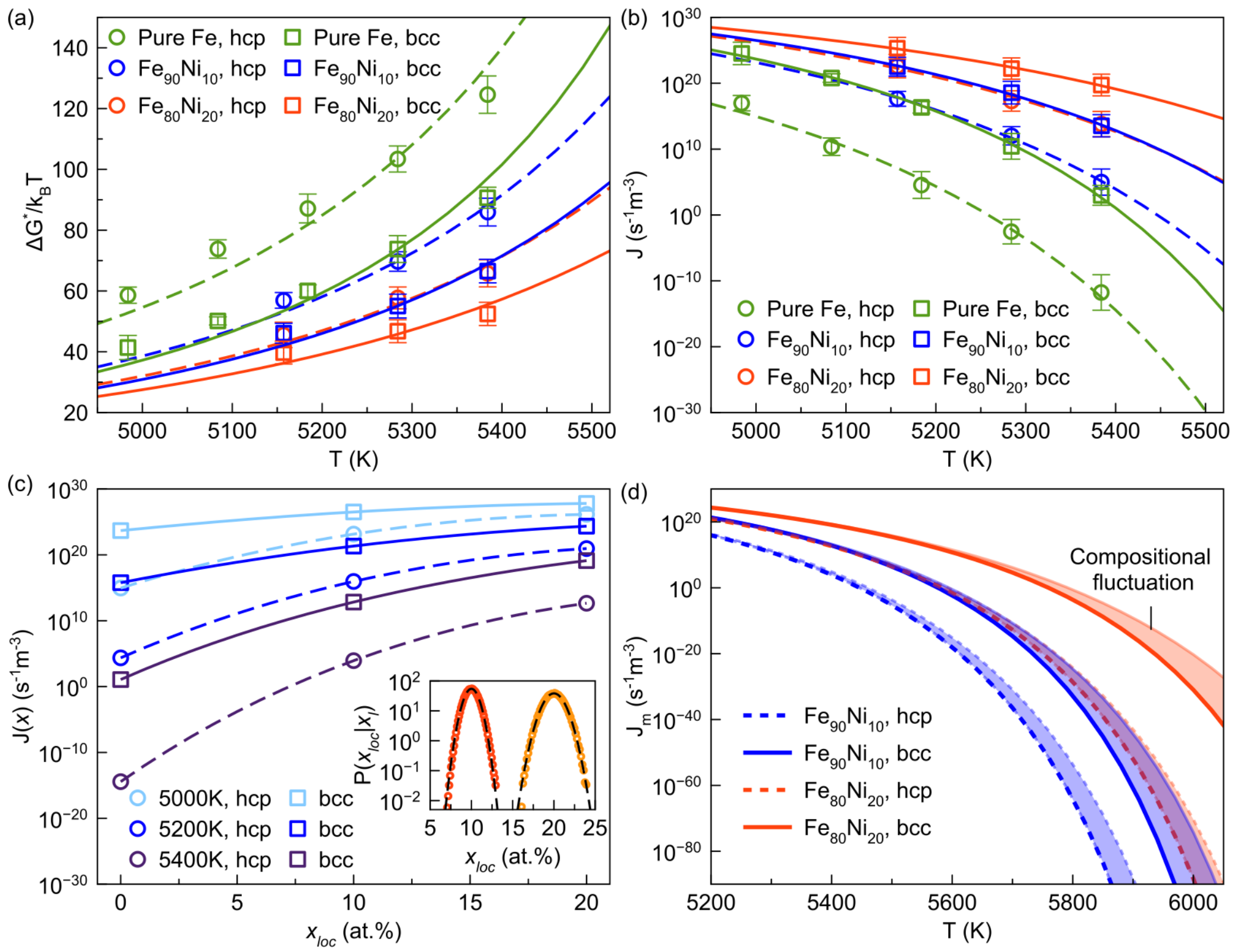

**FIG. 3.** Nucleation barrier and nucleation rate of Fe-Ni alloys. (a) Nucleation barrier versus temperature for pure Fe (green), $Fe_{90}Ni_{10}$ (blue), and $Fe_{80}Ni_{20}$ (red). (b) Nucleation rate versus temperature. Dashed and solid lines are fits to Eq. (2) for the hcp and bcc phases, respectively. (c) Nucleation rate as a function of local Ni composition. Dashed and solid lines show the interpolated $J(x,T)$ for the hcp and bcc phases, respectively. The inset shows the probability distribution of local Ni concentration, $x_{loc}$, for bulk liquids with mean compositions of 10 (red) and 20 (orange) at.% Ni. The dotted lines in the inset are Gaussian fit. (d) Macroscopic nucleation rates including compositional fluctuations. The thin lines with shaded regions show the effect of compositional fluctuations on the nucleation rates relative to the thick lines.

While the nucleus size depends on its composition, the nucleation barrier, $\Delta G^*$, obtained from $\Delta G^* = \frac{1}{2}|\Delta\mu|\, N^*$, is nearly independent of nucleus composition, as shown in Figs. 2(c) and 2(d). Thus, compositional differences in the nucleus primarily affect its size but do not significantly alter the nucleation barrier. Comparison between Figs. 2(c) and 2(d) further indicates that the bcc phase consistently has a lower nucleation barrier than the hcp phase across all temperatures. A similar trend, with composition-dependent nucleus size but composition-independent nucleation barrier, is also observed in the $Fe_{80}Ni_{20}$ liquid, as shown in Supplementary Material Fig. S4. These results indicate that the nucleation barrier in Fe-Ni alloys is primarily controlled by the liquid composition. For instance, the $Fe_{80}Ni_{20}$ liquid exhibits a nucleation barrier approximately 20% lower than that of $Fe_{90}Ni_{10}$ for both hcp and bcc phases. Within the same liquid composition, the trends that decreasing temperature reduces the critical nucleus size and that the bcc phase has a lower nucleation barrier than the hcp phase are consistent with those observed in pure Fe [11,28].

Figure 3(a) and (b) show the nucleation barrier $\Delta G^*$ and rate $J$ computed by $J = \kappa \exp\left(-\frac{\Delta G^*}{k_B T}\right)$, where $\kappa$ is a kinetic prefactor. The bcc phase exhibits a consistently lower barrier than hcp. The addition of Ni significantly reduces the barrier for both bcc and hcp. For instance, at $T = 5400$ K, the hcp nucleation barrier decreases by about 20 $k_B T$ when comparing pure Fe to $Fe_{90}Ni_{10}$. These differences of nucleation barrier translate into orders-of-magnitude variations in the nucleation rate $J$, as illustrated in Fig. 3(b). Higher Ni composition and formation of the bcc phase dramatically increase the nucleation rate. For instance, at $T = 5400$ K, the fastest rate of bcc nucleation in the $Fe_{80}Ni_{20}$ liquid is about $10^{35}$ times higher than hcp nucleation rate in pure Fe liquid. For all nucleation-rate data with the same composition $x$ in Fig. 3(b), the temperature dependence can be well fitted by the empirical formula [55],

$$J(x,T) = \Gamma(x)\exp\left\{-\frac{B(x)T^2}{[T - T_m(x)]^2}\right\}, \qquad (2)$$

where $T_m(x)$ is the composition-dependent melting temperature, and $\Gamma(x)$ and $B(x)$ are fitting parameters. We further describe the compositional dependence of the fitted $\Gamma(x)$ and $B(x)$ using quadratic functions in Supplementary Material Table S1, which provides an interpolation of $J(x,T)$. The resulting $J(x,T)$ curves, plotted in Fig. 3(c), agree well with the calculated raw data.

Given that the nucleation rate is highly sensitive to Ni concentration, it is essential to account for compositional fluctuations in the liquid arising from thermal fluctuations or structural inhomogeneities. To quantify these fluctuations in a stable liquid without convection, we performed large-scale MD simulations of Fe-Ni alloys containing $2 \times 10^6$ atoms for $5\,ns$. The simulation cell was then divided into $10 \times 10 \times 10$ subdomains, each containing approximately 2000 atoms. The local Ni fraction within each sub-volume is computed to construct the probability density $P(x_{loc} \mid x_l)$. Unlike the asymmetric distribution in Fe–O [33], the local Ni concentration distributions are well described by Gaussian statistics, as shown in the inset of Fig. 3c. By accounting for compositional non-uniformity, the macroscopic nucleation rate can be calculated as a spatial average of local contributions from regions with different compositions, expressed as $\bar{J} = \frac{1}{V}\int J(c)dV(c)$ [56]. By invoking spatial ergodicity, the volume fraction $\frac{1}{V}dV(x_{loc})$ occupied by transient regions with a specific local composition is statistically equivalent to the probability density $P(x_{loc}|x_l)dx_{loc}$ of encountering such a region. Accordingly, we formulate the macroscopic nucleation rate including compositional fluctuations, $J_m$, as

$$J_m = \int_0^1 J(x_{loc})P(x_{loc} \mid x_l)\, dx_{loc} \qquad (3)$$

where $P(x_{loc}|x_l)$ is the probability density of finding a local composition $x_{loc}$ within a bulk liquid characterized by a mean composition $x_l$. $J(x_{loc})$ is the nucleation rate described by Eq. (2), assuming a uniform compositional distribution within the local volume. As shown in Fig. 3(d), incorporating these fluctuations via Eq. (3) systematically increases the nucleation rate by orders of magnitude, particularly at higher temperatures.

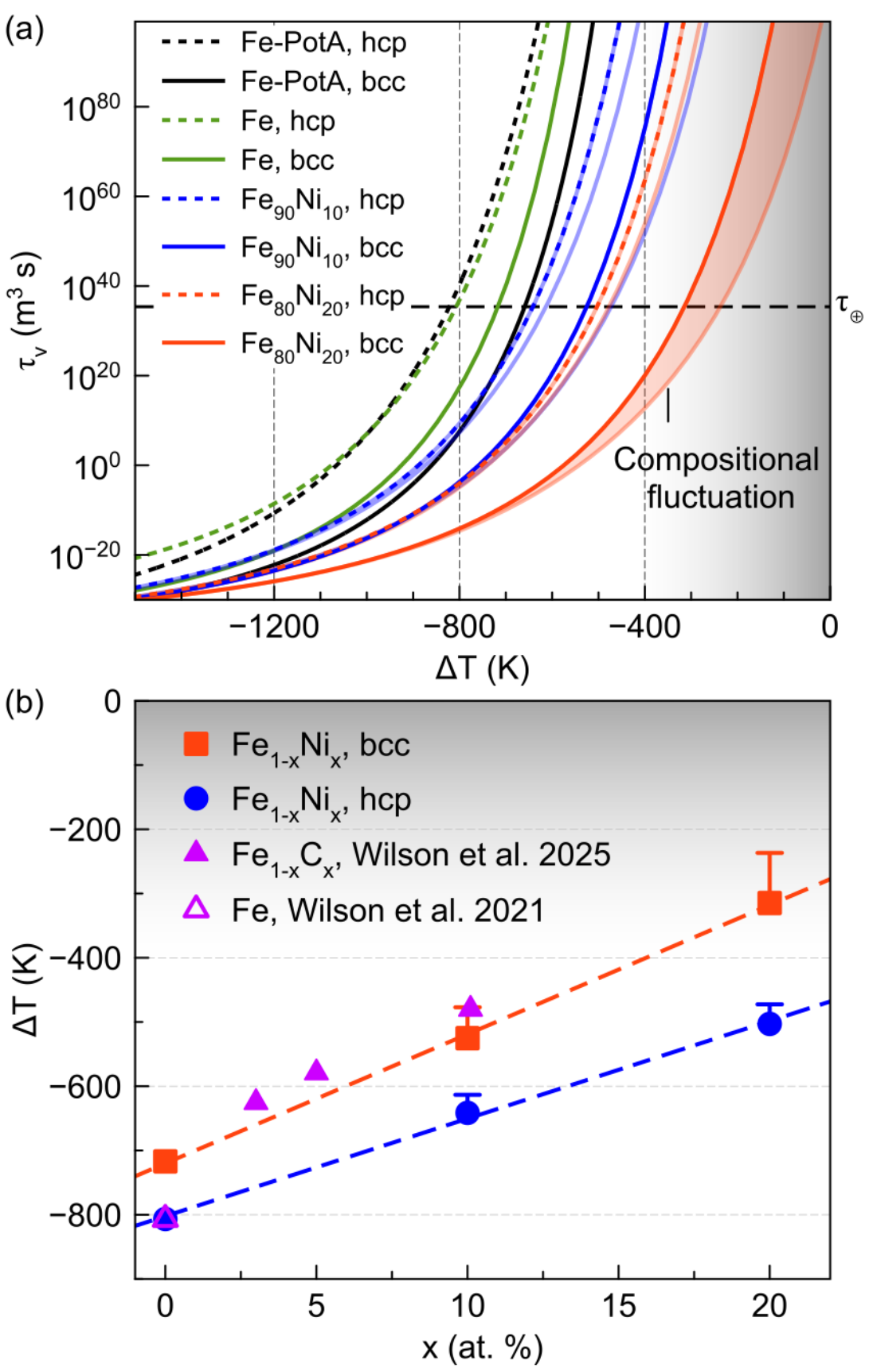

**FIG. 4.** Nucleation waiting time and undercooling for IC formation. (a) Nucleation waiting time, $\tau_v$, versus undercooling relative to the melting point, ΔT, for Fe and Fe-Ni alloys. Dotted and solid lines represent the hcp and bcc phases, respectively. The horizontal dashed line shows the waiting time for the IC nucleation, $\tau_{\oplus}$. The gray shaded region represents the undercooling range for different constraints, with an upper bound of 420 K from [31] and less than 100 K from [9,57]. The Fe-PotA data are the waiting times from [28] computed using Alfè's interatomic potential [58]. The colored shaded curves indicate the waiting time incorporating compositional fluctuations. The thick line shows the data without compositional fluctuations, while the thin lines with shaded regions show the data with compositional fluctuations. (b) Required undercooling to reach $\tau_{\oplus}$ for hcp and bcc nucleation in liquids with different Ni compositions. The bars indicate the effect of compositional fluctuations. Previous data for Fe [59] and $Fe_{1-x}C_x$ [32] are included.

### 3.3 Inner-Core Nucleation

IC nucleation is triggered only after the liquid becomes undercooled below the melting temperature [10]. The allowable range of undercooling can be constrained from several perspectives [9]. A geophysical constraint is inferred from the IC size and core temperature profiles obtained by varying the mineral-physics properties of core materials, giving an upper bound of 420 K [31]. The IC growth model provides another estimate, assuming that a undercooled core would undergo rapid initial freezing, leaving trapped liquid within the IC. Linking the volume of trapped liquid to the initial rapid growth suggests a undercooling of less than 100 K [57,60,61]. In addition, thermal-history models combined with the palaeomagnetic record suggest that supercooling should remain below ~100 K [9]. These constraints are not fully independent, because the inferred nucleation condition, latent heat release, and subsequent IC growth are coupled. Fully resolving the apparent discrepancy between the ~420 K mineral-physics upper bound and the <100 K geophysical estimate requires a comprehensive model coupling core composition, latent heat, and trapped liquid, which is beyond our present scope. We adopt the broader range up to 420 K as a compatible undercooling estimate.

To compare with these undercooling constraints, Figure 4(a) converts the calculated nucleation rates of Fe-Ni alloy into nucleation waiting times, defined as $\tau_v = 1/2J$ [11,30]. Assuming a nucleation incubation time on the order of 1 Gyr and a nucleation volume of ~$7.6\times10^{18}$ $m^3$ i.e., the present-day IC volume, we estimate possible core nucleation waiting time, $\tau_{\oplus}$, to be on the order of $10^{35}$ $m^3\cdot s$, as indicated in Fig. 4. For pure Fe, hcp nucleation requires unrealistically deep undercooling of 805 K to meet this timescale, consistent with the result of [59]. In contrast, the metastable bcc nucleation substantially reduces the waiting time, shifting the required undercooling to smaller magnitudes. This highlights the central kinetic role of bcc formation.

Alloying with Ni further accelerates nucleation. At a given supercooling, both $Fe_{90}Ni_{10}$ and $Fe_{80}Ni_{20}$ exhibit shorter waiting times than pure Fe, and this effect becomes stronger with increasing Ni content. In particular, bcc nucleation in $Fe_{80}Ni_{20}$ liquid shifts the required supercooling close to the <420 K mineral-physics constraint, reflecting the combined effects of bcc formation and reduced nucleation barriers in Ni-rich liquids. Furthermore, incorporating compositional fluctuations in Fig. 4b reduces the required supercooling by another ~100 K, bringing it into the ~250 K regime. Together, these data indicate that bcc formation, Ni-rich melts, and local compositional fluctuations act cooperatively to facilitate IC nucleation.

**4. Discussion**

Our findings support initial bcc nucleation as a kinetic gateway that bypasses the high nucleation barrier for hcp formation. This mechanism was first established for pure Fe, where the metastable bcc phase was found to have a lower nucleation barrier in the melt [11]. Comparative PEM simulations using different interatomic potentials have further confirmed this mechanism under IC conditions [28]. However, quantitative estimates of the waiting time differ among simulations using different interatomic potentials [28]. For instance, PotM, developed by [11], gives a required undercooling of 610 K for hcp nucleation in pure Fe, whereas simulations using PotA, developed by [58], give a required undercooling of ~ 800 K [28]. This difference can be traced to the *ab initio* data used to develop these interatomic potentials. PotM was fitted using the PAW8 potential, which gives an Fe melting temperature of 5858 K at 323 GPa. In contrast, PotA shows an Fe melting temperature of 6215 K at 323 GPa, closer to the *ab initio* value of 6357 K obtained using the more accurate PAW16 potential for describing the electronic structure at the same pressure [20]. The present work addresses this discrepancy by fine-tuning the Fe-Fe interaction in the EAM potential based on the *ab initio* free energy of Fe [62], yielding an Fe melting temperature of 6354 K in close agreement with the PAW16 result. As shown in Fig. 4, the present waiting time for hcp nucleation in pure Fe is therefore consistent with that obtained using PotA, allowing the effect of Ni on Fe nucleation to be compared more directly with PotA-based estimates obtained using nucleation simulations other than PEM [30,31,63]. However, the mixing enthalpy and full liquidus curve in Fig. 1 suggests other deviations from the *ab initio* data arising from the description of Fe–Ni interactions. Addressing this deficiency requires further development of machine-learning potentials [64] to enable nucleation simulations with *ab initio* accuracy.

The present results emphasize the critical role of Ni in IC nucleation. Within the compositional range examined here, Ni concentrations of 5-15 at.% substantially reduces the waiting time and shifts the required supercooling toward the 600–400 K regime, as shown in Fig. 4(b). This range approaches the upper bound of 420 K allowed by the geophysical constraints based on the IC radius [31], although it remains above the more realistic estimate of <100 K [9]. Previous work on Fe-light-element alloys suggested that carbon can also accelerate nucleation, but more than 11 at.% C (i.e. 2.6 wt.%) would be required to achieve nucleation at ~400 K undercooling [31,32]. As compared in Fig. 4(b), C has an effect similar to that of Ni in reducing the required undercooling. However, such a carbon content is much higher than the geochemical estimates of ~0.2 wt.% C in the liquid core from metal-silicate partitioning experiments [65]. Similarly, the Ni content required for nucleation in the present study is also much higher than the mean core Ni concentration of 5.2 wt.% estimated by geochemical models [29]. A coupled effect between Ni and C may provide a possible route to simultaneously satisfying nucleation and geochemical constraints.

Another important factor revealed here is that compositional variations in Fe-Ni liquid can strongly affect nucleation. The local compositional fluctuations considered here arise solely from atomic self-diffusion under equilibrium fluctuations in a chemically homogeneous model and therefore represent the smallest-scale contribution to this effect. In the liquid core, however, compositional heterogeneity is unlikely to be limited to local equilibrium fluctuations. The geomagnetic field has persisted for much of Earth's history, with paleomagnetic records suggesting an active geodynamo as early as the Hadean, implying long-lived convection in the liquid core well before IC nucleation [66]. Such convection could have continually advected and reorganized chemical heterogeneity in the liquid core [67]. Numerical models of compositional convection show that chemical plumes and blobs can persist over long timescales because compositional anomalies diffuse much more slowly than thermal anomalies [68]. Recent seismic observations further suggest that the outer core may be far from well mixed on decadal timescales, with localized transient flows that may be chemically enriched [69]. These processes could broaden the distribution of local compositions sampled by nucleation events beyond the equilibrium fluctuations captured in our atomistic simulations.

Chemical stratification provides another geophysically important source of compositional heterogeneity. Recent thermodynamic studies show that chemical potential equilibrium, including the effect of gravity, can produce radial gradients in light-element concentrations, causing chemical distributions to differ substantially from their bulk averages in the core [70,71]. Although these studies mainly concern light elements rather than Ni, they demonstrate that the liquid core may not be compositionally uniform at length scales relevant to core evolution. In such a chemically heterogeneous liquid, stratification and convection could modify the background composition sampled by nucleation events, while atomic diffusion broadens the local composition around this background. Because the nucleation rate is highly sensitive to composition, these processes may increase the probability of transient Ni-rich or otherwise chemically favorable regions, further reducing the required supercooling and promoting initial IC nucleation. Therefore, during IC formation, the liquid core may have sampled a broad and dynamically evolving compositional field. A natural next step toward addressing the paradox of IC nucleation is to couple Fe-Ni-light-element crystallization kinetics with long-term models of chemical heterogeneity. Moreover, heterogeneous nucleation may reduce the nucleation barrier, but it remains difficult to propose plausible scenarios for a preexisting stable substrate in the early core. Searching for refractory alloys within multicomponent core-forming materials can therefore be important for exploring this route.

In summary, we quantified how Ni affects Fe nucleation under Earth's inner-core conditions by combining free-energy calculations with persistent-embryo simulations. We find that inner-core nucleation is favored by a metastable bcc pathway and that increasing Ni content further lowers the nucleation barrier and reduces the required undercooling. When compositional fluctuations are included, bcc nucleation in $Fe_{80}Ni_{20}$ reaches the ~250 K supercooling regime, substantially narrowing the gap between atomistic nucleation kinetics and geophysical constraints on inner-core formation. These results suggest that Ni enrichment and chemical heterogeneity are important ingredients in resolving the inner-core nucleation paradox.

**Acknowledgements**

Work at Xiamen University was supported by the National Natural Science Foundation of China (Nos. 42374108, T2422016, T25B2015). The supercomputing time was supported by the IKKEM Intelligent Computing Center.